\documentclass[12pt]{article}

\catcode`\@=11
\@addtoreset{equation}{section}

\global\arraycolsep=2pt
\usepackage{amsbsy,amssymb,latexsym,amsfonts,amsmath}
\usepackage{graphicx,color}

\allowdisplaybreaks

\begin{document}
\begin{flushright}
\parbox{4.2cm}
{}
\end{flushright}

\vspace*{0.7cm}

\begin{center}
{\Large \bf 
4D and 2D superconformal index with surface operator}
\vspace*{2.0cm}\\
{Yu Nakayama}
\end{center}
\vspace*{-0.2cm}
\begin{center}
{\it  California Institute of Technology, Pasadena, CA 91125, USA
}
\vspace{3.8cm}
\end{center}

\begin{abstract} 
We study the superconformal index of the $\mathcal{N}=4$ super-Yang-Milles theory on $\mathbf{S}^3 \times \mathbf{S}^1$ with the half BPS superconformal surface operator (defect) inserted at the great circle of $\mathbf{S}^3$. The half BPS superconformal surface operators preserve the same supersymmetry as well as the symmetry of the chemical potential used in the definition of the superconformal index, so the structure and the parameterization of the superconformal index remain unaffected by the presence of the surface operator. On the surface defect, a two-dimensional $(4,4)$ superconformal field theory resides, and the four-dimensional superconformal index may be regarded as a superconformal index of the two-dimensional $(4,4)$ superconformal field theory coupled with the four-dimensional bulk system. We construct the matrix model that computes the superconformal index with the surface operator when it couples with the bulk $\mathcal{N}=4$ super-Yang-Milles theory through the defect hypermultiplets on it.

\end{abstract}

\thispagestyle{empty} 

\setcounter{page}{0}

\newpage

\section{Introduction} 
The index of the four-dimensional superconformal field theory \cite{Romelsberger:2005eg}\cite{Kinney:2005ej} is a superconformal analogue of the Witten index of the supersymmetric quantum mechanics. It is based on the unitary BPZ inner-product of the superconformal field theory on $\mathbf{S}^3 \times \mathbf{R}^1$ rather than the Dirac inner product of quantum mechanics on $\mathbf{R}^4$, and it is invariant under the deformation of the superconformal theory. It captures the BPS spectrum of the superconformal field theory in radial quantization. Because of the invariance, it has been used as a stringent test for various dualities of the four-dimensional superconformal field theories \cite{Nakayama:2005mf}\cite{Nakayama:2006ur}\cite{Romelsberger:2007ec}\cite{Dolan:2008qi}\cite{Spiridonov:2008zr}\cite{Gadde:2009kb}\cite{Spiridonov:2009za}\cite{Gadde:2010te}\cite{Spiridonov:2010hh}\cite{Spiridonov:2010qv}\cite{Vartanov:2010xj}\cite{Gadde:2010en}\cite{Gadde:2011ik}.

It was argued that the superconformal index is the only protected quantity of the superconformal field theories under exactly marginal deformations \cite{Kinney:2005ej}. The argument was based on the fact that the superconformal index is invariant under small changes of the Hilbert space of the theory on $\mathbf{S}^3 \times \mathbf{R}^1$. We may, however, probe the superconformal field theory by drastically changing the structure of the Hilbert space while preserving some of the superconformal symmetry. This is precisely what we would like to do by introducing superconformal defect operators \cite{Constable:2002xt}\cite{Gukov:2006jk}.

In this paper, we would like to study the effect of the superconformal surface operators\footnote{We use the terminology ``surface operator" and ``surface defect" interchangeably.} on the superconformal index. Since the introduction of the surface operators changes the Hilbert space in a discontinuous manner, the superconformal index does change. Yet, the superconformal index captures the BPS spectrum of the superconformal field theory  with the superconformal surface defect inserted. The superconformal index with the superconformal defect is invariant under the exactly marginal deformations of the theory as well as the deformations of the surface operators. As a result, we may compute it either in the weak or strong coupling limit.
This opens up a novel arena of studying the superconformal field theories and their dualities from the superconformal index with superconformal defects.

The half BPS superconformal surface operator preserves the same supersymmetry as well as the symmetry of the chemical potential used in the definition of the superconformal index, so the structure and the parameterization of the superconformal index remain unaffected by the presence of the surface operator. On the surface defect, a two-dimensional $(4,4)$ superconformal field theory resides, and the four-dimensional superconformal index may be regarded as a superconformal index of the two-dimensional $(4,4)$ superconformal field theory coupled with the four-dimensional bulk system.

In particular, we can conjecture that the electric defects and the magnetic defects give the same contribution to the superconformal index as long as they are related by the S-duality simply because the superconformal index is invariant with respect to the change of the gauge coupling constant that is exchanged under teh S-duality. One of the aims of this paper is to provide tools to formulate and understand this claim.

The organization of the paper is as follows. In section 2, we review various facts about the properties of surface operators in $\mathcal{N}=4$ super-Yang-Milles theory on $\mathbf{S}^3 \times \mathbf{R}^1$. In section 3, we define the supreconformal index with surface operators. We give the two-dimensional interpretation of the superconformal index from the defect field theory viewpoint. In section 4,  we construct the matrix model that computes the superconformal index with the surface operator when it couples with the bulk $\mathcal{N}=4$ super-Yang-Milles theory through the defect hypermultiplets on it. In section 5, we further investigate some aspects of the superconformal index with surface operators and conclude.

\section{Superconformal Surface Operators on $\mathbf{S}^3 \times \mathbf{S}^1$}
In this paper we investigate the half BPS superconformal surface operators, which are codimension two defects, of the $\mathcal{N}=4$ super-Yang-Milles theory on $\mathbf{S}^3 \times \mathbf{R}^1$. The codimension two superconformal defects of the $\mathcal{N}=4$ super-Yang-Milles theory on $\mathbf{R}^4 = \mathbf{C}^2$ have been well-studied in the literatures \cite{Constable:2002xt}\cite{Gukov:2006jk}\cite{Gomis:2007fi}\cite{Buchbinder:2007ar}\cite{Drukker:2008wr}\cite{Gukov:2008sn}, and the surface operators that we will study in this paper are all obtained by the conformal transformation from $\mathbf{C}^2$ to $\mathbf{S}^3 \times \mathbf{R}^1$, so
let us begin with the surface operators on $\mathbf{C}^2 = (z_1,z_2)$.

We are interested in the half BPS superconformal defects that preserve 8 of the supersymmetry and 8 of the superconformal symmetry of the $\mathcal{N}=4$ super-Yang-Milles theory on $\mathbf{C}^2$. We put the codimension two defects at $z_2=0$ that preserve the following bosonic symmetry\footnote{We freely perform the Wick rotation when necessary without further notice.}:
\begin{align}
SO(2,4) \times SU(4) \to SL(2,\mathbf{R}) \times SL(2,\mathbf{R}) \times U(1)_{23} \times SU(2)_L \times SU(2)_R \times U(1)_{45} \ .
\end{align}
Here, $SL(2,\mathbf{R}) \times SL(2,\mathbf{R})\in SO(2,4) $ naturally acts on the $z_2=0$ plane as the two-dimensional global conformal transformation (conventionally denoted by $L_0, L_{\pm 1}$ and $\bar{L}, \bar{L}_{\pm 1}$), and the extra $U(1)_{23} \in SO(2,4)$ comes from the rotation of the surface operator in the $z_2$-plane.  The original $SU(4)$ R-symmetry is broken down to $SU(2)_L \times SU(2)_R\times U(1)_{45}$.

A typical example of the surface operators that preserve the above symmetry is the one studied in \cite{Gukov:2006jk} as a higher dimensional analogue of the Wilson- 't Hooft loop operators. For each Cartan subalgebra of the gauge group, we may introduce the magnetic parameter $\alpha$ that allows the singularity of the gauge field
\begin{align}
A = \alpha d \theta \ ,
\end{align}
where $z_2 = r e^{i\theta}$, 
and the electric parameter $\eta$ that gives an additional phase 
\begin{align}
\exp (i\eta \int_{z_2=0} F) \ ,
\end{align}
in the path integral. We can also generalize the surface operator by incorporating the scale invariant configuration of the adjoint Higgs field $(\Phi \sim 1/z_2)$.\footnote{The introduction of the Higgs field $\Phi \sim 1/z_2$ does break the $U(1)_{23} -U(1)_{45}$. In the most part of the paper, the breaking is irrelevant because they do not appear in the superconformal algebra relevant for our study. Only the unbroken combination $U(1)_{23} + U(1)_{45}$ is important.}  The moduli space of the surface operators is given by that of the solutions of the Hitchin's equation with a specific boundary condition and gives rise to a hyper-K\"aher manifold.

Another example of the surface operator is given by the intersecting D3-brane defect studied in  \cite{Constable:2002xt}. We begin with putting $N$ D3-brane (which yields the bulk $\mathcal{N}=4$ super-Yang-Milles) in the $(0,1,2,3)$ direction of the flat ten-dimensional Minkowski space in the type IIB string theory. Then we put another set of the probe D3$'$-branes in the $(0,1,4,5)$ direction. The D3-branes and D3$'$-branes are intersecting at the origin of $(2,3)$- and $(4,5)$-plane that gives the surface defect in the first $\mathcal{N}=4$ super-Yang-Milles theory (from D3-D3 string) in the decoupling limit of the second $\mathcal{N}=4$ super-Yang-Milles theory (from D$3'$-D$3'$ string). The D3-D3$'$ strings yield the localized degree of freedom introducing  a defect (bifundamental) hypermultiplet with $(4,4)$ supersymmetry.

On the superconformal surface defect, a two-dimensional $(4,4)$ superconformal field theory resides. In the limit when the two-dimensional superconformal field theory decouples from the four-dimensional $\mathcal{N}=4$ super-Yang-Milles theory, it must show the full Virasoro symmetry as well as an affine Kac-Moody symmetry, realizing the infinite dimensional $(4,4)$ superconformal algebra (in the NS-NS sector). However, the coupling to the bulk degrees of freedom breaks the full Virasoro symmetry as well as the affine Kac-Moody symmetry. With the bulk degrees of freedome, the theory preserves only the global part of the $(4,4)$ superconformal algebra $SL(2,\mathbf{R}) \times SL(2,\mathbf{R}) \times SU(2)_L \times SU(2)_R$ as well as an additional non-chiral $U(1)$ current $U(1)_\mathcal{J} = U(1)_{23} + U(1)_{45}$, which gives a central extension of the superconformal algebra.

Out of 16 supercharges $Q_{\alpha}^I$ and $Q_{\dot{\alpha}}^I$, where $\alpha, \dot{\alpha} = 1,2$ are spinor indices and $I = 1,\dots,4$ are $SU(4)$ R-symmetry index, the half BPS surface operator preserves $Q_{1}^A$ with $A=1,2$ and $Q_{2}^{A'}$ with $A'=3,4$ (similarly with the hermitian conjugate $Q_{\dot{1} A}$ and $Q_{\dot{2} {A'}}$ on $\mathbf{R}^4$). It also preserves half of the superconformal charges $S_{2B}$ with $B=1,2$  and $S_{1B'}$ with $B'=3,4$ (as well as their hermitian conjugate $S_{\dot{1}}^B$ and $S_{\dot{2}}^{B'}$ on $\mathbf{R}^4$)  out of the 16 superconformal charges $S_{\alpha,J}$ and $S_{\dot{\alpha},J}$ of the $\mathcal{N}=4$ super-Yang-Milles theory. The most relevant piece of the  anti-commutation relation for our discussion is\footnote{We follow the convention used in \cite{Kinney:2005ej}. See Appendix A there.}
\begin{align}
\{S_{\alpha_I}, Q^{\beta J}\} = \delta^{J}_I \delta^{\beta}_\alpha \frac{H}{2} + \delta_{I}^J (J_1)^\beta_{\alpha} + \delta^{\beta}_\alpha R^{J}_I \ .
\end{align}
If we take one particular pair of $Q = Q^{2 1}$ and $S = S_{21}$, we have
\begin{align}
2\{S, Q \} = E -2j_1 - \frac{3}{2} R_1 - R_2 -\frac{1}{2} R_3 \ , \label{mosti}
\end{align}
where  $E$ is the conformal dimension (radial energy), $j_1$ is the angular momentum, and
$R_k$ denotes three Cartan subgroups of $SO(6)$ in the $SU(4)$ notation.

Since the surface operator considered here is invariant under the special conformal transformation acting on the $z_2 = 0$ plane, it is immediate to obtain the superconformal surface operator on $\mathbf{S}^3 \times \mathbf{R}^1$ by a conformal transformation from $\mathbf{C}^2$ to $\mathbf{S}^3 \times \mathbf{R}^1$.  After the conformal transformation, the $z_2 = 0$ plane is located at the great circle of $\mathbf{S}^3$ and along the radial time direction $\mathbf{R}^1$. Furthermore, we will compactify the radial time direction to $\mathbf{S}^1$ in order to define the superconformal index as we will do in the next section. On $\mathbf{S}^3 \times \mathbf{R}^1$, the anti-commutation relation \eqref{mosti} can be understood as the anti-commutation relation between the supercharge $\mathcal{Q}$ and its BPZ conjugate $\mathcal{Q}^\dagger$:
\begin{align}
2\{\mathcal{Q}^\dagger,\mathcal{Q}\} = E -2j_1 - \frac{3}{2} R_1 - R_2 -\frac{1}{2} R_3 \ge 0 \ . \label{anti}
\end{align}
The last inequality is due to the unitarity of the BPZ inner-product (or mathematically known as the Shapovalov form \cite{Sha}). Note that the anti-commutation relation \eqref{anti} was the starting point to define the superconformal index for the $\mathcal{N}=4$ super-Yang-Milles theory on $\mathbf{S}^3 \times \mathbf{S}^1$, and so it is with the surface operator as we will see.

Under the same conformal transformation, the two-dimensional $(4,4)$ superconformal field theory living on the surface defect at $z_2=0$ is mapped to the superconformal field theory on the cylinder $\mathbf{S}^1 \times \mathbf{R}$. The anti-commutation relation \eqref{anti} will be understood as a two-dimensional BPS bound of conformal dimensions \cite{Constable:2002xt}:\footnote{We note that even with the breaking of $U_{23}(1)- U_{45}(1)$ due to the Higgs field profile, the structure of the two-dimensional superconformal algebra is intact.}
\begin{align}
2 \{G^{++}_{1/2}, G^{--}_{-1/2}\} =  2h - 2j^3_L + \mathcal{J} \ge 0 \ .  \label{tda}
\end{align}
On the left hand side, we have used the conventional notation for the $\mathcal{N}=4$ superconformal algebra where $G^{++}_{1/2}$ is the left-moving supercharge with the left-moving conformal dimension $1/2$ and the R-charge $+1$, and  $G^{--}_{-1/2}$ is the left-moving supercharge with the left-moving conformal dimension $-1/2$ and the R-charge $-1$.
On the right hand side, $h$ is the left-moving conformal dimension, $j^3_L$ is the left-moving $SU(2)$ R-charge, and $\mathcal{J} = J_{23} + J_{45}$ is the central extension. 
Actually, the same $\mathcal{J}$ appears in the ``right-moving" BPS algebra
\begin{align}
2 \{\bar{G}^{++}_{1/2}, \bar{G}^{--}_{-1/2}\} =  2\bar{h} - 2{j}^3_R + \mathcal{J} \ge 0 \ ,  \label{tdas}
\end{align}
where $\bar{h}$ is the right-moving conformal dimension, and $j^3_R$ is the right-moving $SU(2)$ R-charge. The the appearance of the $\mathcal{J}$ shows the non-decoupling of the left-mover and right-mover.
If we restrict ourselves to the $\mathcal{J}=0$ sector, the anti-commutation relation \eqref{tda} is precisely the chiral primary condition of the $(4,4)$ superconformal algebra. As we will see, localized states on the defect show $\mathcal{J} = 0$.

For future reference, we summarize the dictionary between the quantum numbers of the four-dimension and their counterparts in two-dimension:
\begin{align}
E = h + \bar{h} \ , \ \ j_1 = -\frac{1}{2}(h-\bar{h} + J_{23})  \ , \ \ j_2 = -\frac{1}{2}(h - \bar{h} - J_{23})  \ , \cr
R_1 = 2j_L^3 \ , \ \ R_2 = -(j_L^3 + j_R^3 + J_{45}) , \ \ R_3 = 2 j^3_R \ , \ \ \mathcal{J} = J_{23} + J_{45} \ . 
\end{align}

\section{Superconformal Index with Surface Operator}
The superconformal index of the $\mathcal{N}=4$ super-Yang-Milles theory \cite{Kinney:2005ej} is defined by
\begin{align}
\mathcal{I}(t,y,v,w) = \mathrm{Tr} (-1)^F e^{-\beta \Delta} t^{2(E+j_1)} y^{2j_2}v^{R_2} w^{R_3} \ , \label{indexd}
\end{align}
where $\Delta = 2\{\mathcal{Q}^\dagger, \mathcal{Q}\} = E -2j_1 - \frac{3}{2} R_1 - R_2 -\frac{1}{2} R_3 $. The trace is taken over the Hilbert space on $\mathbf{S}^3 \times \mathbf{R}^1$.
Due to the bose-fermi cancellation, the superconformal index does not depend on the radial temperature $\beta$, and only the states with $\Delta = 0$ will contribute. The chemical potentials $t,y,v,w$ are chosen so that the corresponding charges $E+j_1, j_2, R_2,R_3$ all commute with the supercharge $\mathcal{Q}$.
The superconformal index is invariant under the marginal deformation of the theory, and in particular it is known that the superconformal index does not show a wall-crossing (at least in the large $N$ limit) so that the superconformal index is independent of the Yang-Milles coupling constant as well as the $\theta$-parameter \cite{Kinney:2005ej}. 

One can evaluate the superconformal index by using the localization technique to directly compute the path integral \cite{Kinney:2005ej}\cite{Nawata:2011un} or by simply counting the gauge invariant local operators  \cite{Kinney:2005ej} (see also \cite{Bianchi:2006ti} for the group theoretical derivation). In either way, the computation of the superconformal index for $U(N)$ gauge group reduces to the effective matrix integral \cite{Kinney:2005ej}:
\begin{align}
\mathcal{I}(t,y,v,w) &\equiv \mathrm{Tr} (-1)^F t^{2(E+j_1)} y^{2j_2} v^{R_2} w^{R_3}\cr
&=\int [dU] e^{-S_{\mathrm{eff}}[U]} \ , \label{matrix}
\end{align}
where the effective matrix action is given by
\begin{align}
&-S_{\mathrm{eff}}[U] =\cr
&\sum_{n>0} \frac{1}{n}\frac{t^{2n}(v^n+w^{-n}+\frac{w^{n}}{v^{n}})-t^{3n}(y^n+y^{-n})-t^{4n}(w^n+v^{-n}+\frac{v^n}{w^{n}})+2t^{6n}}{(1-y^{n}t^{3n})(1-y^{-n}t^{3n})}\chi_a(U^n) \  \label{effmat}
\end{align} 
with the adjoint character:  $\chi_a(U^n) = \mathrm{Tr}U^n\mathrm{Tr} U^{-n}$. 
The integration is over the  unitary matrix $U$ with the invariant Haar measure $[dU]$. The physical interpretation of the matrix $U$ is that it is the Polyakov loop $ U = P\exp(\oint A_0 dt)$ along the radial time.

The exact integration is possible in the $N\to \infty$ limit by using the saddle point approximation valid in the large $N$ matrix integral, which yields an elegant  expression for  $U(N)$ gauge group \cite{Kinney:2005ej}: 
\begin{eqnarray}
\mathcal{I}_{U(\infty)} = \prod_{n>0} \frac{(1-t^{3n} y^n)(1-t^{3n}y^{-n})}{(1-t^{2n}/w^{n})(1-t^{2n}w^n/v^n)(1-t^{2n} v^n)} \ . \label{indexr}
\end{eqnarray}

The superconformal index shows an interesting limiting structure as noted in \cite{Nakayama:2007jy}. For later purpose, we briefly discuss the three-dimensional limit studied in \cite{Dolan:2011rp}\cite{Gadde:2011ia}\cite{Imamura:2011uw}\cite{Nishioka:2011dq}, where we reduce the $\mathcal{N}=4$ super-Yang-Milles theory on $\mathbf{S}^3 \times \mathbf{S}^1$ to the $\mathcal{N}=8$ super-Yang-Milles theory on $\mathbf{S}^3$ by shrinking the radial time circle. By following the same reasoning in \cite{Gadde:2011ia}\cite{Imamura:2011uw},  the limit is taken by setting $y=1$, $t = v = e^{-r/3}$ and $w= e^{-r/6}$ with $r \to 0$, where $r$ plays the role of the effective radius of the Kaluza-Klein circle. The limit must accompany the zeta-function regularization before the integration over the holonomy matrix $U$, and we cannot take the limit directly in \eqref{indexr}, but we should take the limit in \eqref{matrix}. The resultant reduced matrix model formally agrees with the zeta-function regularized supersymmetric partition function of the $\mathcal{N}=8$ super-Yang-Milles theory on $\mathbf{S}^3$ obtained by a naive application of the localization technique in \cite{Kapustin:2009kz}\cite{Kapustin:2010xq}, where ``naive" meaning that the infrared R-symmetry assignment used in the localization is incorrect.\footnote{The difficulty is that the ``correct" R-charge assignment is not manifest in the ultraviolet Lagrangian but it is an emerging one in the infrared limit \cite{Kapustin:2010xq}.}

With this regard, we should note that the naive application of the localization to the $\mathcal{N}=8$ super-Yang-Milles theory leads to the physically unacceptable result, and as a result, we should take the naive three-dimensional limit from the superconformal index with a grain of salt. Indeed one can easily see that the resulting holonomy integral never converges and does not give a meaningful result. Nevertheless, we will pursue the three-dimensional limit further in section 5.3 with the insertion of the surface operator, hoping that the same technique must apply in less supersymmetric situations with the correct R-charge assignment.

Now we would like to define the superconformal index of the $\mathcal{N}=4$ super-Yang-Milles theory with the half BPS superconformal surface operator inserted along the great circle of $\mathbf{S}^3$ and the radial time circle. As we have discussed in the last section, the superconformal surface operator preserves the same supersymmetry $\mathcal{Q}$ and its BPZ conjugate $\mathcal{Q}^\dagger$ that was used in the definition of the superconformal index. In addition, all the charges associated with the chemical potential $E + j_1$, $j_2$, $R_2$, and $R_3$ are preserved by the existence of the half BPS superconformal surface operator. Therefore, we can recycle the same definition of the superconformal index 
\begin{align}
\mathcal{I}_S(t,y,v,w) = \mathrm{Tr}_S (-1)^F e^{-\beta \Delta} t^{2(E+j_1)} y^{2j_2}v^{R_2} w^{R_3} \ , \label{indexs}
\end{align}
where the trace is now taken over the Hilbert space of the $\mathcal{N}=4$ super-Yang-Milles theory coupled with the superconformal defect placed at the great circle of $\mathbf{S}^3$ and wrapped around the radial time direction. 

Again due to the bose-fermi cancellation, the superconformal index does not depend on the radial temperature $\beta$ and only the states with $\Delta = 0$ will contribute. Since the necessary conformal transformation is not broken by the surface operator, one can use the familiar state-operator correspondence, and regard the superconformal index with the surface operator as counting of the BPS local gauge invariant operators, which satisfy the condition $\Delta = 0$, of the $\mathcal{N}=4$ super-Yang-Milles theory on $\mathbf{C}^2$ coupled with the half BPS superconformal surface operator placed at $z_2=0$.

We have sufficient knowledge about the local gauge invariant operators coming from the $\mathcal{N}=4$ super-Yang-Milles theory, so the novel contributions to the superconformal index mainly come from the state localized on the surface defect. From the defect field theory viewpoint, it is more natural to treat the superconformal index with the surface operator from the two-dimensional field theory perspective. As discussed in the last section, the half BPS superconformal surface operator hosts a two-dimensional (4,4) superconformal field theory. Furthermore, we can also reduce the four-dimensional bulk field on the surface defect with the additional ``Kaluza-Klein" parameter ``$z_2$" integrated over (details of the reduction can be found in \cite{Constable:2002xt}).

To appreciate the two-dimensional structure, we perform a change of variables for the chemical potential, and we may well regard the four-dimensional superconformal index (with the surface operator) as a two-dimensional superconformal ``index"
\begin{align}
\mathcal{I}_S(\bar{q},\bar{z},\bar{y},\bar{w}) = \mathrm{Tr}_{\mathrm{2D;NSNS}} (-1)^F q^{\Delta} \bar{q}^{\bar{L}_0} \bar{z}^{J_R^3} \bar{y}^{2j_2} \bar{w}^{R_2} \ ,
\end{align}
where $ t = \bar{q}^{1/4}$, $y = \bar{q}^{1/4} \bar{y}$, $v = \bar{z}$ and $w = \bar{w}$.
The relation between the four-dimensional charge and the two-dimensional charge is $2\bar{L}_0 = E + j_1 + j_2$, which is the ``right-moving" Virasoro energy, and $J_R^3 = R_3$, which is the ``right-moving" $SU(2)$ R-symmetry of the $(4,4)$ superconformal algebra.\footnote{Strictly speaking, the complete separation of the ``left-mover" and the ``right-mover" requires the decoupling of the bulk $\mathcal{N}=4$ super-Yang-Milles theory, but the global charges are nevertheless separately conserved without the decoupling.}
Of course, in the above expression, only the $\Delta =0$ state will contribute to the superconformal index, so it is independent of $q$ (that couples with the (twisted) ``left-moving" Virasoro energy).

We can turn off the ``extra" chemical potential $\bar{y}$ and $\bar{w}$, which would not appear in vanilla $(4,4)$ superconformal algebra, by setting $\bar{y}=\bar{w}=1$ to make it intrinsic to two-dimension\footnote{The interpretation of the other two chemical potential is as follows: the one is $\mathcal{J}$ which is zero for the localized operators on the defect, and the other involves the charge which could be broken by the profile of the Higgs field as we have discussed in footnote 3 and 5.} as $\mathrm{Tr}_{\mathrm{2D;NSNS}} (-1)^F q^{\Delta} \bar{q}^{\bar{L}_0} \bar{z}^{J_R^3}$. In the decoupling limit of the interaction between the bulk $\mathcal{N}=4$ super-Yang-Milles theory and the defect $(4,4)$ superconformal field theory, only the ``left chiral operators" that satisfy $L_0 = j^L_3$ contribute to the superconformal index from the defect $(4,4)$ superconformal field theory side. This is because the simple relation $\Delta = 2L_0 - 2 j_3^L$ holds for the operators with $\mathcal{J} =0$ that are localized on the defect. This superconformal ``index" is an NS-NS analogue of the elliptic genera of the $(4,4)$ SCFT conventionally defined in the R-R sector. If one decouples the bulk super-Yang-Milles theory degree of freedom, the $(4,4)$ SCFT living on the surface operator is typically given by the sigma model whose target space is hyper-K\"ahler (from the supersymmetry), so one may compute the superconformal index directly within the two-dimensional conformal field theory. In particular, we may expect a nice modular property with respect to the parameters $\bar{q}$ and $\bar{z}$ (see section 4 for a concrete example).

The coupling with the bulk $\mathcal{N}=4$ super-Yang-Milles theory degree of freedom, however, will introduce the non-zero $\mathcal{J}$ sector to the Hilbert state. Recall that $\mathcal{J} = J_{23}+J_{45} = j_2 -j_1 -\frac{R_1}{2}-R_2-\frac{R_3}{2}$, and the $\mathcal{N}=4$ super-Yang-Milles fields are charged under $\mathcal{J}$ (see table 1). As we mentioned in the last section, after coupling with the bulk, the infinite Virasoro symmetry and Kac-Moody symmetry can no longer be present precisely due to the existence of non-zero $\mathcal{J}$. As a consequence, we do not expect any nice modular property of the full superconformal index. Indeed, as an extreme example, one may consider the insertion of the  ``trivial" surface operator, which gives the $\mathcal{N}=4$ superconformal index itself, but we have not been aware of any nice modular property of such even after the change of variables although the $N \to \infty$ result \eqref{indexr} may look slightly promising.\footnote{The exchange of the great circle of $\mathbf{S}^3$ and the radial time is not an isometry of the system unlike the simple $\mathbf{S}^1 \times \mathbf{S}^1$ where the defect degrees of freedom live, so we do not expect that a nice modular property would exist.}

The superconformal index with the surface operator is naturally protected against the continuous deformation of the theory as long as it preserves the superconformal symmetry. In particular, we expect that it would not change over the moduli space of the superconformal surface operators studied e.g. in \cite{Gukov:2006jk}. It should be interesting to see if and how it jumps when the gauge symmetry breaking pattern of the surface operators (so-called Levi group) changes at the singular point of the moduli space.

\section{Matrix Model for Superconformal Index with Defect Hypermultiplet}
So far, we have discussed the general features of the superconformal index with  the superconformal surface operator. In this section, we would like to compute the superconformal index with the surface operator in the simplest example when the surface operator is given by the defect $(4,4)$ hypermultiplet superconformal field theory. This is physically realized by the intersecting D3-brane model studied in \cite{Constable:2002xt}.

As in the bulk superconformal index for the $\mathcal{N}=4$ super-Yang-Milles theory, we can either use the localization technique to evaluate the path integral directly, or by counting the gauge invariant local operators of the theory with the superconformal surface defect. Since the same supercharge relevant for the localization of the bulk $\mathcal{N}=4$ super-Yang-Milles theory is preserved under the insertion of the superconformal surface operator, we can use the same localization procedure. The computation must reduce to the matrix integral over the  Polyakov loop $U = P \exp(\oint A_0 dt)$. 

In this section, we study the counting problem of the defect $(4,4)$ hypermultiplet superconformal field theory coupled with the bulk $\mathcal{N}=4$ super-Yang-Milles theory on $\mathbf{C}^2$ with the defect at $z_2=0$. As a concrete model, we discuss the intersecting D3-brane defect studied in \cite{Constable:2002xt}. In their setup, the total Lagrangian (on $\mathbf{C}^2$) is the sum of the bulk $\mathcal{N}=4$ super-Yang-Milles theory and the defect $(4,4)$ hypermultiplet. The defect $(4,4)$ hypermultiplet is charged under the bulk $\mathcal{N}=4$ super-Yang-Milles theory, and this gauging provides the coupling between the bulk and the defect. Without the gauging, the defect $(4,4)$ hypermultiplet forms the trivial hyper K\"ahler structure (simply by a tensor product of $\mathbf{R}^4$). The free part of the defect action is given by
\begin{align}
L = \int d^2z \left(-|Db|^2 -|D\tilde{b}|^2 + i\bar{\psi}_b^- D \psi_b^- + i \bar{\psi}_b^+ \bar{D} \psi_b^+ +  i\bar{\psi}_{\tilde{b}}^- D \psi_{\tilde{b}}^- + i \bar{\psi}_{\tilde{b}}^+ \bar{D} \psi_{\tilde{b}}^+ \right)  \ .
\end{align}
We can find the interacting action with the bulk $\mathcal{N}=4$ super-Yang-Milles theory in Appendix D of \cite{Constable:2002xt}, but except for the fact that the defect hypermultiplet transforms under the gauge symmetry, the precise form of the interaction is irrelevant for our study of the index.

\begin{table}[tb]
\begin{center}
\begin{tabular}{c|c|c|c|c}
 Letters        & $(-1)^F[E,j_1,j_2] $& $[R_1,R_2,R_3]$ & rep & index \\
 \hline
   $X,Y,Z$&     $  [1,0,0] $         &$[0,1,0],[1,-1,1],[1,0,-1]$ & adj & $t^2(v+\frac{w}{v}+\frac{1}{w})$ \\
  $\bar{{\psi}}_X, \bar{\psi}_Y, \bar{\psi}_Z $&     $  -[\frac{3}{2},\frac{1}{2},0] $ &$[1,-1,0],[0,1,-1],[0,0,1] $&adj & $-t^4(\frac{1}{v}+\frac{v}{w} +w)$ \\
\hline
$F_{++}$&     $ [2,1,0] $ &$0 $&adj &$t^6$ \\
${\lambda}_{\pm} $&     $ -[\frac{3}{2},0,\pm\frac{1}{2}] $ &$[1,0,0] $&adj& $-t^3(y+\frac{1}{y})$ \\
\hline
$ \partial_\mu \sigma^\mu \lambda =0  $&     $ [\frac{5}{2},\frac{1}{2},0] $ &$[1,0,0] $&adj &$t^6$ \\
\hline
$\partial_{+\pm} $&     $ [1,\frac{1}{2},\pm\frac{1}{2}] $ &$0 $&1 &$t^3 y, t^3y^{-1}$   
 \end{tabular}
\end{center}
\caption{The letters that will contribute to the single particle index from the bulk $\mathcal{N}=4$ super-Yang-Milles theory.}
\label{tab:1}
\end{table}%

\begin{table}[tb]
\begin{center}
\begin{tabular}{c|c|c|c|c}
 Letters        & $(-1)^F[E,j_1,j_2] $& $[R_1,R_2,R_3]$ & rep & index\\
 \hline
 $b$&     $  [0,\frac{1}{4},-\frac{1}{4}] $         &$[0,-\frac{1}{2},0]$ & $ R $  &  $t^{\frac{1}{2}}y^{-\frac{1}{2}}v^{-\frac{1}{2}}$ \\
  $\tilde{b} $&     $  [0,\frac{1}{4},-\frac{1}{4}] $&$[0,-\frac{1}{2},0] $ & ${R}^*$ &$t^{\frac{1}{2}}y^{-\frac{1}{2}}v^{-\frac{1}{2}} $\\
 $b^*$&     $  [0,-\frac{1}{4},\frac{1}{4}] $         &$[0,\frac{1}{2},0]$ & $ {R}^* $  &  $t^{-\frac{1}{2}}y^{\frac{1}{2}}v^{\frac{1}{2}}$ \\
  $\tilde{b}^* $&     $  [0,-\frac{1}{4},\frac{1}{4}] $&$[0,\frac{1}{2},0] $ & ${R}$ &$t^{-\frac{1}{2}}y^{\frac{1}{2}}v^{\frac{1}{2}} $\\
  \hline
   $\psi_b^+$&     $  -[\frac{1}{2},-\frac{1}{4},-\frac{1}{4}] $         &$[1,-\frac{1}{2},0]$ & $R$& $-t^{\frac{1}{2}}y^{-\frac{1}{2}} v^{-\frac{1}{2}}$ \\
  $\bar{\psi}_{\tilde{b}}^+$&     $  -[\frac{1}{2},-\frac{1}{4},-\frac{1}{4}] $         &$[-1,\frac{1}{2},0]$ & $R$& NA \\  
$\bar{\psi}_b^+$&     $  -[\frac{1}{2},-\frac{1}{4},-\frac{1}{4}] $         &$[-1,\frac{1}{2},0]$ & $R^*$& NA \\
  $\psi_{\tilde{b}}^+$&     $  -[\frac{1}{2},-\frac{1}{4},-\frac{1}{4}] $         &$[1,-\frac{1}{2},0]$ & $R^*$& $-t^{\frac{1}{2}}y^{-\frac{1}{2}} v^{-\frac{1}{2}}$ \\
  \hline
 $\psi_b^-$&     $  -[\frac{1}{2},\frac{1}{4},\frac{1}{4}] $         &$[0,-\frac{1}{2},1]$ & $R$& $-t^{\frac{3}{2}}y^{\frac{1}{2}} v^{-\frac{1}{2}} w$ \\
  $\bar{\psi}_{\tilde{b}}^-$&     $  -[\frac{1}{2},\frac{1}{4},\frac{1}{4}] $         &$[0,\frac{1}{2},-1]$ & $R$& $-t^{\frac{3}{2}} y^{\frac{1}{2}} v^{\frac{1}{2}} w^{-1}$ \\  
$\bar{\psi}_b^-$&     $  -[\frac{1}{2},\frac{1}{4},\frac{1}{4}] $         &$[0,\frac{1}{2},-1]$ & $R^*$&$-t^{\frac{3}{2}} y^{\frac{1}{2}} v^{\frac{1}{2}} w^{-1}$ \\
  $\psi_{\tilde{b}}^-$&     $  -[\frac{1}{2},\frac{1}{4},\frac{1}{4}] $         &$[0,-\frac{1}{2},1]$ & $R^*$& $-t^{\frac{3}{2}}y^{\frac{1}{2}} v^{-\frac{1}{2}}w$ \\
  \hline
$\partial_{z} $&     $ [1,-\frac{1}{2},-\frac{1}{2}] $ &$0 $&$1$ & NA \cr
$\partial_{\bar{z}} $& $[1, \frac{1}{2},\frac{1}{2}] $ & $0$&$1$ & $t^{3} y$ \\ 
 \end{tabular}
\end{center}
\caption{The letters of the defect hypermultiplet on the superconformal surface operator. The index with ``NA" means that they do not contribute to the superconformal index because they do not satisfy $\Delta = 0$. We have to impose the dirac equation $\partial_{\bar{z}} \psi^+ = 0$ on $\psi^+_b$ and $\psi^+_{\tilde{b}}$ to correctly count the left-moving fermionic degrees of freedom. For the normalizability of $E=0$ states, see the main text.}
\label{tab:2}
\end{table}%

The counting of the single particle letter of the bulk $\mathcal{N}=4$ super-Yang-Mills theory that satisfies the BPS condition $\Delta = 0$ was done in \cite{Kinney:2005ej} and we simply quote their results together with the single particle index \eqref{indexd} in table 1. The counting of the single particle letter of the defect $(4,4)$ hypermultiplet is also straightforward. We present the results together with the single particle index \eqref{indexs} in table 2. 

There is a small subtlety in counting the BPS states and computing the superconformal index with the superconformal surface operator. The problem is that the two-dimensional field theory with a massless scalar suffers an infrared divergence. The appearance of the infrared divergence is related to the fact that we have $E=0$ scalar (seemingly BPS) operators $b$ and $\tilde{b}$ in the defect hypermultiplet. The corresponding states are non-normalizable, and the two-point functions among these operators do not scale with power laws: they show logarithmic tales. The usual prescription is to declare that only the operators with the derivative, say $\partial_{\bar{z}} b$, consist of the normalizable Hilbert state of the massless scalar theory in two-dimension. From the path integral viewpoint, we divide the partition function by the volume of the constant zero-mode of the sigma model target space (which is infinite in our example). 

In the following discussion, we exclude the logarithmically non-normalizable state to compute the superconformal index. Indeed, if we did not throw away these non-normalizable contributions, the superconformal index would not converge because positive as well as negative powers of $t$ appear (actually infinitely many times) in the formal expression of the superconformal index. Thus, in comparison with table 2, the lowest bosonic operators that contribute to the infrared divergence free superconformal index begin with $\partial_{\bar{z}}b, \partial_{\bar{z}}\tilde{b} ,\partial_{\bar{z}}b^*$ and $\partial_{\bar{z}}\tilde{b}^*$  rather than  the naked $b$ or $\tilde{b}$. These normalizable operators possess the letter index $ t^{\frac{7}{2}} y^{\frac{1}{2}} v^{-\frac{1}{2}}$ and  $t^{\frac{5}{2}} y^{\frac{3}{2}} v^{\frac{1}{2}}$ rather than $t^{\frac{1}{2}} y^{-\frac{1}{2}}v^{-\frac{1}{2}}$ and $t^{-\frac{1}{2}}y^{\frac{1}{2}} v^{\frac{1}{2}}$. Note that with the derivative, the contribution to the superconformal index always has a positive power of $t$, ensuring the convergence of the superconformal index.

With these remarks in mind, we now show how the computation of the superconformal index reduce to the matrix model integral. 
To compute the superconformal index, we take the plethystic exponential of the single particle letter and integrate it over the holonomy of the gauge group in order to project it down to gauge singlet states.  The integration over the holonomy $U$ has again a nice interpretation of the localized path integral over the flat connection (i.e. Polyakov loop along the radial time) on $\mathbf{S}^3  \times \mathbf{S}^1 $.

From table 2, one can easily read the contribution of the defect hypermultiplet to the superconformal index.
The effective matrix action from one defect hypermultiplet is given by
\begin{align}
 &-S_{\mathrm{eff;s}}[U] = \sum_{n>0} \left[ \frac{1}{n}  \right. \cr
& \left(\frac{t^{\frac{5n}{2}} y^{\frac{3n}{2}} v^{\frac{n}{2}} + t^{\frac{7n}{2}} y^{\frac{n}{2}} v^{-\frac{n}{2}} -t^{\frac{3n}{2}}y^{\frac{n}{2}}v^{-\frac{n}{2}} w^{n} -  t^{\frac{3n}{2}}y^{\frac{n}{2}}v^{\frac{n}{2}} w^{-n}}{1-y^n t^{3n}} -t^{\frac{n}{2}}y^{-\frac{n}{2}} v^{-\frac{n}{2}} \right) \chi_s(U^n)  \cr
 +&  \left. \left(\frac{t^{\frac{5n}{2}} y^{\frac{3n}{2}} v^{\frac{n}{2}} + t^{\frac{7n}{2}} y^{\frac{n}{2}} v^{-\frac{n}{2}} -t^{\frac{3n}{2}}y^{\frac{n}{2}}v^{-\frac{n}{2}} w^{n} -  t^{\frac{3n}{2}}y^{\frac{n}{2}}v^{\frac{n}{2}} w^{-n}}{1-y^n t^{3n}}  -t^{\frac{n}{2}}y^{-\frac{n}{2}} v^{-\frac{n}{2}} \right) \chi_s (U^{-n})   \right]  \ . \label{defect}
\end{align}
The character $\chi_s(U^n)$ depends on the representation of the hypermultiplet. For instance,  the fundamental representation has $\chi_s(U^n) = \mathrm{Tr} U^n$ and $\chi_s(U^{-n}) = \mathrm{Tr} U^{-n}$. 

The entire superconformal  index incorporating the contribution from the bulk $\mathcal{N}=4$ super-Yang-Milles theory is computed by the matrix integral
\begin{align}
\mathcal{I}(t,y,v,w) =\int [dU] e^{-S_{\mathrm{eff;bulk}}[U] - S_{\mathrm{eff;s}}[U]} \ . 
\end{align}
The bulk part of the effective action is given by \eqref{effmat} as before. When the hypermultiplet is in the fundamental representation, the explicit integral over the holonomy is difficult. In this case, we are not aware of a simple large $N$ technique to solve the matrix integral because it does not seem to reduce to the Gaussian integral with the fundamental defect hypermultiplet. When the defect hypermultiplet is in the adjoint representation, we may perform the integral in the large $N$ limit as in \cite{Kinney:2005ej}.

For reference, we show the first few terms of the superconformal index that comes from one defect hypermultiplet in the fundamental representation for $U(N)$ gauge group ($N >1$):
\begin{align}
 \mathcal{I}_S(t,y,v,w) &= 1 + t y^{-1} v^{-1} + 3t^2 v^{-1} w + 3 t^2 w^{-1} + t^2 v - t^3 y + t^3y^{-1} \cr
&+ t^3 y v^{-1}w^2 + t^3 y v w^{-2} + 2t^3 y^{-1} w v^{-2} + 2t^3 y^{-1} v^{-1} w^{-1} +  \mathcal{O}(t^4) \ .
\end{align}
It agrees with the brute-force counting of gauge invariant operators. 
For example, $\psi_b \psi^+_{\tilde{b}}$ gives $t y^{-1}v^{-1}$. $\psi^+_b \bar{\psi}_{b}^-$, $\psi^+_{\tilde{b}} \bar{\psi}^-_{\tilde{b}}$ and $\mathrm{Tr} Y$ give $3t^2 w^{-1}$. $\psi^+_{b} \psi^{-}_{\tilde{b}}$, $\psi^+_{\tilde{b}} \psi^-_{b}$ and $\mathrm{Tr} Z$ give $3t^2 v^{-1} w$, and $\mathrm{Tr} X$ gives $t^2v$. $\partial_{\bar{z}} b^* \psi_b^+$, $\partial_{\bar{z}} \tilde{b}^* \psi^+_{\tilde{b}}$, $\psi_b^- \bar{\psi}^-_b$, $\psi_{\tilde{b}}^- \bar{\psi}^-_{\tilde{b}}$, and $\mathrm{Tr}\lambda_{+}$ gives $-t^3y$ and so on.

In the last section, we mentioned a possible modular property of the superconformal index with the surface operator. As discussed there, there seems no theoretical evidence why it should show any interesting modular property unless we decouple the bulk $\mathcal{N}=4$ super-Yang-Milles theory degree of freedom. If we decouple the bulk $\mathcal{N}=4$ super-Yang-Milles theory and do not impose the gauge singlet condition on the operators localized on the defect, we may expect an interesting modular property with the change of variables suitable for the two-dimensional interpretation (see section 3).

In our example, the two-dimensional ``index"
\begin{align}
\mathcal{I}_S(q,z,y,w)&= \mathcal{I}_S(\bar{q},\bar{z},\bar{y},\bar{w}) \cr
&= \mathrm{Tr}_{\mathrm{2D;NSNS}} (-1)^F q^{\Delta} \bar{q}^{\bar{L}_0} \bar{z}^{J_R^3} \bar{y}^{2j_2} \bar{w}^{R_2} \ 
\end{align}
actually vanishes by setting $\bar{y} = \bar{w} =1$ after the above-mentioned decoupling. This is due to the fact that the left-mover has four chiral primary states (or R-vacua after the spectral flow): $1$, $\psi_b^+$, $\psi_{\tilde{b}}^+$ and $\psi_b^+ \psi_{\tilde{b}}^+$, and they cancel with each other. Note that Hilbert space of the free field theory considered here is the direct product of the decoupled left-mover and right-mover.
If we artificially neglected these left-moving ``fermionic zero-mode" contribution, the rest of the index from the right-mover would be just given by the partition function of two chiral bosons and two chiral NS-fermions (for each hypermultiplet) explicitly given by $\bar{q}^{1/4}\theta_{01}^2(\bar{q},\bar{w})/\eta^6(\bar{q})$ and would show a conventional modular property that exchanges the NS-fermions with the periodic boundary condition and the R-fermions with the anti-periodic boundary condition.

\section{Further Discussions and Conclusion}
\subsection{AdS/CFT}

The $\mathcal{N}=4$ super-Yang-Milles theory in large $N$ limit enjoys the AdS/CFT correspondence. The computation of the superconformal index of the $\mathcal{N}=4$ super-Yang-Milles theory on $\mathbf{S}^3 \times \mathbf{S}^1$ from the supergravity has been performed in \cite{Kinney:2005ej} and showed the complete agreement with the weak coupling computation from the gauge theory side. This confirms that the superconformal index is indeed invariant under the change of the gauge coupling constant and does not show the wall-crossing.

The gravity dual description of the superconformal surface operators for $\mathcal{N}=4$ super-Yang-Milles theory have been investigated in \cite{Gomis:2007fi}\cite{Buchbinder:2007ar}\cite{Drukker:2008wr}. Again, the AdS/CFT has been successful in understanding the behavior of the superconformal surface operators. The surface operators are understood as a probe D3-brane in the AdS space or bubbling supergravity solution depending on the class of surface operators considered.

For our applications, we note that the spectrum of the gravity dual of the  intersecting D3-brane system was studied in \cite{Constable:2002xt}. They showed that the bosonic fluctuation of the D3-brane probe in the AdS space completely agrees with the BPS spectrum of the defect $(4,4)$ superconformal field theory. Since we are counting the same BPS states in the computation of the index with the intersecting surface defect, their agreements imply that the AdS/CFT computation of the index must be possible.

A small subtlety is that in their comparison, they included the logarithmically non-normalizable modes. As we have showed,  once the logarithmically non-normalizable modes are allowed, we encounter the severe infrared divergence and the index does not converge. Since they included a certain restricted class of non-normalizable modes in their comparison, it would be interesting to see how their restriction can be made precise in our index computation. Leaving aside this subtlety,  our computation is completely consistent with their analysis in the bosonic sector.

\subsection{Less supersymmetry}
In this paper, we have studied the index of the $\mathcal{N}=4$ super-Yang-Milles theory with surface operator inserted. Most of the discussions in this paper is applicable to less supersymmetric cases. For instance, the $\mathcal{N}=2$ superconformal field theory admits surface operators preserving the bosonic symmetry \cite{Alday:2009fs}\cite{Gaiotto:2009fs}:
\begin{align}
SL(2,\mathbf{R}) \times SL(2,\mathbf{R}) \times U(1)_L \times U(1)_R \times U(1)_{\mathcal{J}} \in SO(2,4) \times SU(2)_R \times U(1)_r
\end{align}
where $SL(2,\mathbf{R}) \times SL(2,\mathbf{R}) \times U(1)_L \times U(1)_R$ will be identified with the bosonic subgroup of the $(2,2)$ superconformal algebra (in NS-NS sector), and the additional $U(1)_{\mathcal{J}}$ (denoted by $U(1)_e$ in  \cite{Alday:2009fs}) plays the role of the non-chiral coupling between the bulk and the defect.

As in the $\mathcal{N}=4$ case, the preserved (super)symmetry is compatible with the index of $\mathcal{N}=2$ superconformal field theories:
\begin{align}
\mathcal{I}(t,y,v) = \mathrm{Tr} (-1)^F e^{-\beta \Delta} t^{2(E+j_1)} y^{2j_2} v^{-\frac{r}{2} - \frac{R}{2}} \ ,
\end{align}
where $\Delta = E-2j_2-R-\frac{r}{2}$. We can define the $\mathcal{N}=2$ superconformal index with the superconformal surface operator by using the same expression with additional contribution from the defect sector. In complete parallel with the $\mathcal{N}=4$ case, the BPS condition $\Delta \ge 0$ of the four-dimensional index is interpreted as the BPS condition of the $(2,2)$ superconformal field theory on the defect:
\begin{align}
2h - 2j_L  + \mathcal{J} \ge 0 \ ,
\end{align}
where $j_L$ is the left-moving $U(1)_R$ charge.
In particular, for the $\mathcal{J}=0$ states localized on the surface defect, the BPS condition is nothing but the chiral primary condition of the $(2,2)$ two-dimensional superconformal field theory.

In the literatures (e.g.  \cite{Alday:2009fs}\cite{Gaiotto:2009fs}), examples of supersymmetric surface operators in $\mathcal{N}=2$ gauge theory have been investigated. They are all classically conformally invariant, but most of them are not conformally invariant quantum mechanically because the effective field theory living on the defect becomes massive. 
Our discussion requires the exact conformal invariance at the quantum level, so we should be careful about the breaking of the conformal invariance. 

In \cite{Gadde:2011ik}, it has been shown that the bulk superconformal index of the $\mathcal{N}=2$ gauge theories are related to supersymmetric partition function of the $q$-deformed two-dimensional Yang-Milles theory and more generically certain topological field theories on Riemann surfaces. It would be interesting to see how we can interpret our index of the $\mathcal{N}=2$ gauge theories with surface operators in terms of the language of the two-dimensional Yang-Milles theory.

We can further reduce the supersymmetry down to $\mathcal{N}=1$ in four-dimension. The superconformal index can be defined in the similar manner. An example of the superconformal surface operator in the Klebanov-Witten theory was studied in \cite{Koh:2009cj}. Note that in the $\mathcal{N}=1$ case, the corresponding two-dimensional $(1,1)$ superconformal algebra does not possess the R-symmetry, so the BPS bound in the two-dimensional interpretation is entirely supported by the $U(1)_{\mathcal{J}}$ symmetry.

\subsection{$\mathbf{S}^3$ reduction}
As discussed in \cite{Dolan:2011rp}\cite{Gadde:2011ia}\cite{Imamura:2011uw}, the four-dimensional superconformal index on $\mathbf{S^3} \times \mathbf{S}^1$ has a very interesting limit, where we reproduce the three-dimensional supersymmetric partition function on $\mathbf{S}^3$.\footnote{The reduction in the $\mathcal{N}=1$ case is anomalous \cite{Nak}\cite{Imamura:2011uw}, so the corresponding four-dimensional index which we would like to take the limit  is not well-defined while the problem is circumvented in $\mathcal{N}=2,4$ case.} With this picture, the four-dimensional index, in particular the integrand of the holonomy integration, can be seen as a $q$-deformation of the three-dimensional partition function on $\mathbf{S}^3$. A mathematics behind is a profound limiting structure of the elliptic hypergeometric functions (see e.g. \cite{elliptic1}\cite{elliptic2}).

The incorporation of the surface operator in the four-dimensional superconformal index naturally leads to the introduction of the loop operators at the great circle of $\mathbf{S}^3$ in the  three-dimensional supersymmetric field theory. Indeed, the logic was precisely the same. The reason why we can compute the three-dimensional supersymmetric partition function with the BPS loop operator as in the same way we compute the partition function without the loop operator is that the both system preserves the same supersymmetry relevant for the localization \cite{Kapustin:2009kz}. We have seen the parallel situation in the four-dimensional index, where the surface operator preserves the same supersymmetry relevant for the definition as well as the computation of the superconformal index.

Unfortunately, it is slightly moot to study the direct three-dimensional limit of our superconformal index with the surface operator inserted for the $\mathcal{N}=4$ super-Yang-Milles theory as it is. This is  because it will turn out to be assigning the wrong R-symmetry in the naive localization, and the holonomy integral will not converge (even without the surface operator: see \cite{Kapustin:2010xq}). The following argument is therefore quite formal, but we hope that the same technique is applicable to the $\mathcal{N}=2$ case, where the problem of the wrong R-charge assignment can be avoided.  In particular, one may study the mirror dual of the $\mathcal{N}=8$ super-Yang-Milles theory by starting with the four-dimensional $\mathcal{N}=2$ theory with fundamental hypermultiplets \cite{Kapustin:2010xq}.

At the formal level, we can demonstrate how to take the three-dimensional limit in our example studied in section 4. By setting $y = 1$, $t=v=e^{-r/3}$ and $w = e^{-r/6}$ with $r\to 0$ within the holonomy integral, we can compute the three-dimensional supersymmetric partition function on $\mathbf{S}^3$ with the conventional but incorrect R-charge assignment.\footnote{One may compute the partition function on squashed $\mathbf{S}^3$ by introducing the more general limit $y = e^{-rs}$, where $s$ is related to the squashing parameter \cite{Dolan:2011rp}\cite{Gadde:2011ia}\cite{Imamura:2011uw}.} In this limit, the bulk $\mathcal{N}=4$ super-Yang-Milles part \eqref{defect} formally reduces to the contribution of the $\mathcal{N}=8$ super-Yang-Milles theory to the three-dimensional supersymmetric partition function (but with a wrong R-charge assignment). On the other hand, the surface operator contribution \eqref{effmat} reduces to the contribution from the line defect in the three-dimensional super-Yang-Milles theory. 

The limit gives the factor $\exp\left(\sum_n \frac{c}{n} (\chi_s(U^n) + \chi_s(U^{-n}))\right)$, where $c$ is a numerical factor that depends on the squashing parameter (for $s=0$, it is $5/3$).  By expanding the exponential, we see that the intersecting defect gives rise to the insertion of the three-dimensional Wilson-loop with various tensor product representations. Of course, this is quite formal in the $\mathcal{N}=8$ super-Yang-Milles theory case because the remaining holonomy integral is divergent in any way, but the structure must remain the same if we consider the reduction of the $\mathcal{N}=2$ gauge theories with surface operators.   

It must be of very importance to see if we could find a dictionary between the superconformal surface operators in $\mathcal{N}=2$ gauge theory in four-dimension and the supersymmetric loop operators of the $\mathcal{N}=4$ supersymmetric field theories in three-dimension through the computation of the index. Note that the superconformal index (with the surface operator) is invariant under the change of the coupling constant of the theory. Thus, the objects that are related by the S-duality must show the {\it same} contribution to the index when the gauge theory is self-dual (like $U(N)$ gauge theory) in four-dimension, and so must be case  also after the reduction to the three-dimension.\footnote{The S-duality non-trivially act on the supercharges as a phase \cite{Kapustin:2006pk}, but the phase is irrelevant for our study of the superconformal index.}   We leave this conjecture for a future study.

\subsection{Loop operators}
Yet another interesting object one can introduce in the superconformal index (with or without the surface defect) is a loop operator along the radial time direction located at a point on $\mathbf{S}^3$. The simplest example is the insertion of the Polyakov loop. With the Polyakov loop, the matrix model is simply modified by the insertion of the matrix character $\chi_r(U)$ within the matrix integral:
\begin{align}
\mathcal{I}_r(t,y,v,w) = \int [dU] \chi_r(U) e^{-S_{\mathrm{eff}}} \ . \label{elect}
\end{align}
For instance, for the fundamental representation, $\chi_r(U) = \mathrm{Tr} U$, and it vanishes (without the surface defect). This is physically expected because the adjoint valued field in $\mathcal{N}=4$ super-Yang-Milles theory cannot form a gauge invariant state with the single heavy spectator fundamental field inserted from the Polyakov loop.

Again, the index does not depend on the coupling constant, so we can exchange the electric defect with the magnetic defect without changing the index as long as the theory is self-dual by S-duality.\footnote{Even if the theory is not self-dual, we have predictions: for instance the index with electric objects in $SP(2N)$ theory must be identical to the index with magnetic objects in $SO(2N+1)$ theory and vice versa.} 
 With the same reasoning, we can argue that the index with the supersymmetric defect insertion only depends on the S-duality orbit of the superconformal objects. It would be very interesting to verify this conjecture by directly studying the path integral of the $\mathcal{N}=4$ super-Yang-Milles theory on $\mathbf{S}^3 \times \mathbf{S}^1$ with the superconformal magnetic defects in comparison with the electric expression \eqref{elect}. 

\subsection{Conclusion}
In this paper, we have studied the superconformal index of the $\mathcal{N}=4$ super-Yang-Milles theory with half BPS superconformal defect. Although our main emphasis is the formalism,  we have constructed the matrix model that computes the superconformal index with the surface operator when it couples with the bulk $\mathcal{N}=4$ super-Yang-Milles theory through the defect hypermultiplets on it.

One of the significant features of the superconformal index is that it is invariant under the marginal deformation of the theory. In particular, it must be invariant under the change of the gauge coupling constant. We, therefore, conjecture that the superconformal index with a superconformal defect is invariant under the S-duality transformation: the superconformal index with the defect operator only depends on the S-duality orbit of the defect. 

In this paper, the computation of the superconformal index is mainly done by counting gauge invariant operators. This picture is particularly suitable for electric defects. It would be interesting to perform the direct path integral by using  localization to compute the index with magnetic defects to verify the conjecture stated in the last paragraph.

\section*{Acknowledgements}
 The work of Y.~N. is supported by Sherman Fairchild Senior Research Fellowship at California Institute of Technology.


\begin{thebibliography}{99}
%\cite{Romelsberger:2005eg}
\bibitem{Romelsberger:2005eg}
  C.~Romelsberger,
  %``Counting chiral primaries in N = 1, d=4 superconformal field theories,''
  Nucl.\ Phys.\  B {\bf 747}, 329 (2006)
  [arXiv:hep-th/0510060].
  %%CITATION = NUPHA,B747,329;%%
%\cite{Kinney:2005ej}
\bibitem{Kinney:2005ej}
  J.~Kinney, J.~M.~Maldacena, S.~Minwalla and S.~Raju,
  %``An Index for 4 dimensional super conformal theories,''
  Commun.\ Math.\ Phys.\  {\bf 275}, 209 (2007)
  [arXiv:hep-th/0510251].
  %%CITATION = CMPHA,275,209;%%

%\cite{Nakayama:2005mf}
\bibitem{Nakayama:2005mf}
  Y.~Nakayama,
  %``Index for orbifold quiver gauge theories,''
  Phys.\ Lett.\  B {\bf 636}, 132 (2006)
  [arXiv:hep-th/0512280].
  %%CITATION = PHLTA,B636,132;%%

%\cite{Nakayama:2006ur}
\bibitem{Nakayama:2006ur}
  Y.~Nakayama,
  %``Index for supergravity on AdS(5) x T**1,1 and conifold gauge theory,''
  Nucl.\ Phys.\  B {\bf 755}, 295 (2006)
  [arXiv:hep-th/0602284].
  %%CITATION = NUPHA,B755,295;%%



%\cite{Romelsberger:2007ec}
\bibitem{Romelsberger:2007ec}
  C.~Romelsberger,
  %``Calculating the Superconformal Index and Seiberg Duality,''
  arXiv:0707.3702 [hep-th].
  %%CITATION = ARXIV:0707.3702;%%

%\cite{Dolan:2008qi}
\bibitem{Dolan:2008qi}
  F.~A.~Dolan and H.~Osborn,
  %``Applications of the Superconformal Index for Protected Operators and
  %q-Hypergeometric Identities to N=1 Dual Theories,''
  Nucl.\ Phys.\  B {\bf 818}, 137 (2009)
  [arXiv:0801.4947 [hep-th]].
  %%CITATION = NUPHA,B818,137;%%

%\cite{Spiridonov:2008zr}
\bibitem{Spiridonov:2008zr}
  V.~P.~Spiridonov and G.~S.~Vartanov,
  %``Superconformal indices for N = 1 theories with multiple duals,''
  Nucl.\ Phys.\  B {\bf 824}, 192 (2010)
  [arXiv:0811.1909 [hep-th]].
  %%CITATION = NUPHA,B824,192;%%
%\cite{Gadde:2009kb}
\bibitem{Gadde:2009kb}
  A.~Gadde, E.~Pomoni, L.~Rastelli, S.~S.~Razamat,
  %``S-duality and 2d Topological QFT,''
  JHEP {\bf 1003}, 032 (2010).
  [arXiv:0910.2225 [hep-th]].
%\cite{Spiridonov:2009za}
\bibitem{Spiridonov:2009za}
  V.~P.~Spiridonov and G.~S.~Vartanov,
  %``Elliptic hypergeometry of supersymmetric dualities,''
  Commun.\ Math.\ Phys.\  {\bf 304}, 797 (2011)
  [arXiv:0910.5944 [hep-th]].
  %%CITATION = CMPHA,304,797;%%


%\cite{Gadde:2010te}
\bibitem{Gadde:2010te}
  A.~Gadde, L.~Rastelli, S.~S.~Razamat, W.~Yan,
  %``The Superconformal Index of the E_6 SCFT,''
  JHEP {\bf 1008}, 107 (2010).
  [arXiv:1003.4244 [hep-th]].

%\cite{Spiridonov:2010hh}
\bibitem{Spiridonov:2010hh}
  V.~P.~Spiridonov and G.~S.~Vartanov,
  %``Supersymmetric dualities beyond the conformal window,''
  Phys.\ Rev.\ Lett.\  {\bf 105}, 061603 (2010)
  [arXiv:1003.6109 [hep-th]].
  %%CITATION = PRLTA,105,061603;%%

%\cite{Spiridonov:2010qv}
\bibitem{Spiridonov:2010qv}
  V.~P.~Spiridonov and G.~S.~Vartanov,
  %``Superconformal indices of ${\mathcal N}=4$ SYM field theories,''
  arXiv:1005.4196 [hep-th].
  %%CITATION = ARXIV:1005.4196;%%



%\cite{Vartanov:2010xj}
\bibitem{Vartanov:2010xj}
  G.~S.~Vartanov,
  %``On the ISS model of dynamical SUSY breaking,''
  Phys.\ Lett.\  B {\bf 696}, 288 (2011)
  [arXiv:1009.2153 [hep-th]].
  %%CITATION = PHLTA,B696,288;%%



%\cite{Gadde:2010en}
\bibitem{Gadde:2010en}
  A.~Gadde, L.~Rastelli, S.~S.~Razamat and W.~Yan,
  %``On the Superconformal Index of N=1 IR Fixed Points: A Holographic Check,''
  JHEP {\bf 1103} (2011) 041
  [arXiv:1011.5278 [hep-th]].
  %%CITATION = JHEPA,1103,041;%%
%\cite{Gadde:2011ik}
\bibitem{Gadde:2011ik}
  A.~Gadde, L.~Rastelli, S.~S.~Razamat, W.~Yan,
  %``The 4d Superconformal Index from q-deformed 2d Yang-Mills,''
  [arXiv:1104.3850 [hep-th]].

%\cite{Constable:2002xt}
\bibitem{Constable:2002xt}
  N.~R.~Constable, J.~Erdmenger, Z.~Guralnik, I.~Kirsch,
  %``Intersecting D-3 branes and holography,''
  Phys.\ Rev.\  {\bf D68}, 106007 (2003).
  [hep-th/0211222].

%\cite{Gukov:2006jk}
\bibitem{Gukov:2006jk}
  S.~Gukov, E.~Witten,
  %``Gauge Theory, Ramification, And The Geometric Langlands Program,''
  [hep-th/0612073].

%\cite{Gomis:2007fi}
\bibitem{Gomis:2007fi}
  J.~Gomis and S.~Matsuura,
  %``Bubbling surface operators and S-duality,''
  JHEP {\bf 0706}, 025 (2007)
  [arXiv:0704.1657 [hep-th]].
  %%CITATION = JHEPA,0706,025;%%
%\cite{Buchbinder:2007ar}
\bibitem{Buchbinder:2007ar}
  E.~I.~Buchbinder, J.~Gomis and F.~Passerini,
  %``Holographic gauge theories in background fields and surface operators,''
  JHEP {\bf 0712}, 101 (2007)
  [arXiv:0710.5170 [hep-th]].
  %%CITATION = JHEPA,0712,101;%%


%\cite{Drukker:2008wr}
\bibitem{Drukker:2008wr}
  N.~Drukker, J.~Gomis and S.~Matsuura,
  %``Probing N=4 SYM With Surface Operators,''
  JHEP {\bf 0810}, 048 (2008)
  [arXiv:0805.4199 [hep-th]].
  %%CITATION = JHEPA,0810,048;%%


%\cite{Gukov:2008sn}
\bibitem{Gukov:2008sn}
  S.~Gukov and E.~Witten,
  %``Rigid Surface Operators,''
  arXiv:0804.1561 [hep-th].
  %%CITATION = ARXIV:0804.1561;%%

\bibitem{Sha} N. N. Shapovalov, Funct. Anal. Appl. 6 (1972), 307.

%\cite{Nawata:2011un}
\bibitem{Nawata:2011un}
  S.~Nawata,
  %``Localization of N=4 Superconformal Field Theory on S^1 x S^3 and Index,''
  [arXiv:1104.4470 [hep-th]].


%\cite{Bianchi:2006ti}
\bibitem{Bianchi:2006ti}
  M.~Bianchi, F.~A.~Dolan, P.~J.~Heslop, H.~Osborn,
  %``N=4 superconformal characters and partition functions,''
  Nucl.\ Phys.\  {\bf B767}, 163-226 (2007).
  [hep-th/0609179].


%\cite{Nakayama:2007jy}
\bibitem{Nakayama:2007jy}
  Y.~Nakayama,
  %``Finite N index and angular momentum bound from gravity,''
  Gen.\ Rel.\ Grav.\  {\bf 39}, 1625-1638 (2007).
  [hep-th/0701208].
%\cite{Dolan:2011rp}
\bibitem{Dolan:2011rp}
  F.~A.~H.~Dolan, V.~P.~Spiridonov and G.~S.~Vartanov,
  %``From 4d superconformal indices to 3d partition functions,''
  arXiv:1104.1787 [hep-th].
  %%CITATION = ARXIV:1104.1787;%%


%\cite{Gadde:2011ia}
\bibitem{Gadde:2011ia}
  A.~Gadde, W.~Yan,
  %``Reducing the 4d Index to the $S^3$ Partition Function,''
  [arXiv:1104.2592 [hep-th]].


%\cite{Imamura:2011uw}
\bibitem{Imamura:2011uw}
  Y.~Imamura,
  %``Relation between the 4d superconformal index and the S^3 partition function,'' 
[arXiv:1104.4482 [hep-th]].

%\cite{Nishioka:2011dq}
\bibitem{Nishioka:2011dq}
  T.~Nishioka, Y.~Tachikawa, M.~Yamazaki,
  %``3d Partition Function as Overlap of Wavefunctions,''
  [arXiv:1105.4390 [hep-th]].



\bibitem{elliptic1}
V.~P.~Spiridonov, Russ. Math. Surv. 56 185 2001.
\bibitem{elliptic2}
V.~P.~Spirindonov, Algebra i Analiz (St. Petersburg Math. J.) 15 (2003) 161-21
[arXiv:math/0303205]

%\cite{Kapustin:2009kz}
\bibitem{Kapustin:2009kz}
  A.~Kapustin, B.~Willett, I.~Yaakov,
  %``Exact Results for Wilson Loops in Superconformal Chern-Simons Theories with Matter,''
  JHEP {\bf 1003}, 089 (2010).
  [arXiv:0909.4559 [hep-th]].

%\cite{Kapustin:2010xq}
\bibitem{Kapustin:2010xq}
  A.~Kapustin, B.~Willett, I.~Yaakov,
  %``Nonperturbative Tests of Three-Dimensional Dualities,''
  JHEP {\bf 1010}, 013 (2010).
  [arXiv:1003.5694 [hep-th]].

%\cite{Alday:2009fs}
\bibitem{Alday:2009fs}
  L.~F.~Alday, D.~Gaiotto, S.~Gukov, Y.~Tachikawa, H.~Verlinde,
  %``Loop and surface operators in N=2 gauge theory and Liouville modular geometry,''
  JHEP {\bf 1001}, 113 (2010).
  [arXiv:0909.0945 [hep-th]].
%\cite{Gaiotto:2009fs}
\bibitem{Gaiotto:2009fs}
  D.~Gaiotto,
  %``Surface Operators in N = 2 4d Gauge Theories,''
  [arXiv:0911.1316 [hep-th]].

%\cite{Koh:2009cj}
\bibitem{Koh:2009cj}
  E.~Koh, S.~Yamaguchi,
  %``Surface operators in the Klebanov-Witten theory,''
  JHEP {\bf 0906}, 070 (2009).
  [arXiv:0904.1460 [hep-th]].


\bibitem{Nak}
Private communication with Y.~Imamura.

%\cite{Kapustin:2006pk}
\bibitem{Kapustin:2006pk}
  A.~Kapustin and E.~Witten,
  %``Electric-Magnetic Duality And The Geometric Langlands Program,''
  arXiv:hep-th/0604151.
  %%CITATION = HEP-TH/0604151;%%

\end{thebibliography}
\end{document}